\documentstyle[12pt]{article}
\begin{document}
\def\IB{\relax\hbox{$\inbar\kern-.3em{\rm B}$}}
\def\IC{\relax\hbox{$\inbar\kern-.3em{\rm C}$}}
\def\ID{\relax\hbox{$\inbar\kern-.3em{\rm D}$}}
\def\IE{\relax\hbox{$\inbar\kern-.3em{\rm E}$}}
\def\IF{\relax\hbox{$\inbar\kern-.3em{\rm F}$}}
\def\IG{\relax\hbox{$\inbar\kern-.3em{\rm G}$}}
\def\IGa{\relax\hbox{${\rm I}\kern-.18em\Gamma$}}
\def\IH{\relax{\rm I\kern-.18em H}}
\def\IK{\relax{\rm I\kern-.18em K}}
\def\IL{\relax{\rm I\kern-.18em L}}
\def\IP{\relax{\rm I\kern-.18em P}}
\def\IR{\relax{\rm I\kern-.18em R}}
\def\IZ{\relax\ifmmode\mathchoice
{\hbox{\cmss Z\kern-.4em Z}}{\hbox{\cmss Z\kern-.4em Z}}
{\lower.9pt\hbox{\cmsss Z\kern-.4em Z}}
{\lower1.2pt\hbox{\cmsss Z\kern-.4em Z}}\else{\cmss Z\kern-.4em Z}\fi}
\def\ocom{\relax{\raise1.2pt\hbox{$\otimes$}\kern-.5em\lower.1pt\hbox{,}}\,}
\def\CA {{\cal A}}
\def\CB {{\cal B}}
\def\CC {{\cal C}}
\def\CD {{\cal D}}
\def\CE {{\cal E}}
\def\CF {{\cal F}}
\def\CG {{\cal G}}
\def\CH {{\cal H}}
\def\CI {{\cal I}}
\def\CJ {{\cal J}}
\def\CK {{\cal K}}
\def\CL {{\cal L}}
\def\CM {{\cal M}}
\def\CN {{\cal N}}
\def\CO {{\cal O}}
\def\CP {{\cal P}}
\def\CQ {{\cal Q}}
\def\CR {{\cal R}}
\def\CS {{\cal S}}
\def\CT {{\cal T}}
\def\CU {{\cal U}}
\def\CV {{\cal V}}
\def\CW {{\cal W}}
\def\CX {{\cal X}}
\def\CY {{\cal Y}}
\def\CZ {{\cal Z}}

\def\wwp{{\wp}_{\tau}(q)}
\def\gpg{g^{-1} \p g}
\def\pgp{\pb g g^{-1}}
\def\p{\partial}
\def\pb{\bar{\partial}}
\def\ib{{\bar i}}
\def\jb{{\bar j}}

\def\ub{{\bar{u}}}
\def\wb {\bar{w}}
\def\zb {\bar{z}}

\def\codim{{\mathop{\rm codim}}}
\def\cok{{\rm cok}}
\def\rank{{\rm rank}}
\def\coker{{\mathop {\rm coker}}}
\def\diff{{\rm diff}}
\def\Diff{{\rm Diff}}
\def\ib{{\bar i}}
\def\Tr{\rm Tr}
\def\Id{{\rm Id}}
\def\vol{{\rm vol}}
\def\Vol{{\rm Vol}}
\font\manual=manfnt \def\dbend{\lower3.5pt\hbox{\manual\char127}}
\def\danger#1{\smallskip\noindent\rlap\dbend%
\indent{\bf #1}\par\vskip-1.5pt\nobreak}
\def\c{\cdot}
\def\half {{1\over 2}}
\def\ch{{\rm ch}}
\def\Det{{\rm Det}}
\def\DET{{\rm DET}}
\def\Hom{{\rm Hom}}
\def\imp{$\Rightarrow$}
\def\IX{{\bf X}}
\def\neib{neighborhood}
\def\inbar{\,\vrule height1.5ex width.4pt depth0pt}

\def\Lie{{\rm Lie}}
\def\lieg{{\underline{\bf g}}}
\def\lieh{{\underline{\bf h}}}
\def\liet{{\underline{\bf t}}}
\def\liek{{\underline{\bf k}}}
\def\sdtimes{\mathbin{\hbox{\hskip2pt\vrule height 4.1pt depth -.3pt width
.25pt
\hskip-2pt$\times$}}}
\def\clb#1#2#3#4#5#6{\pmatrix{#1 & #2 & #3\cr
#4 & #5 & #6\cr} }

\def\QATOPD#1#2#3#4{{#3 \atopwithdelims#1#2 #4}}
\def\stackunder#1#2{\mathrel{\mathop{#2}\limits_{#1}}}
\def\bea{\begin{eqnarray}}
\def\eea{\end{eqnarray}}
\def\nn{\nonumber}
\def\baselinestretch{1.5}
\def\beq{\begin{equation}}
\def\eeq{\end{equation}}
\def\ba{\beq\new\begin{array}{c}}
\def\ea{\end{array}\eeq}
\def\be{\ba}
\def\ee{\ea}
\def\stackreb#1#2{\mathrel{\mathop{#2}\limits_{#1}}}
\def\Tr{{\rm Tr}}
\def\res{{\rm res}}
\def\f{1\over}
\parskip=0.4em
\makeatletter
\newdimen\normalarrayskip              
\newdimen\minarrayskip                 
\normalarrayskip\baselineskip
\minarrayskip\jot
\newif\ifold             \oldtrue            \def\new{\oldfalse}
\def\arraymode{\ifold\relax\else\displaystyle\fi} 
\def\eqnumphantom{\phantom{(\theequation)}}     
\def\@arrayskip{\ifold\baselineskip\z@\lineskip\z@
     \else
     \baselineskip\minarrayskip\lineskip2\minarrayskip\fi}
\def\@arrayclassz{\ifcase \@lastchclass \@acolampacol \or
\@ampacol \or \or \or \@addamp \or
   \@acolampacol \or \@firstampfalse \@acol \fi
\edef\@preamble{\@preamble
  \ifcase \@chnum
     \hfil$\relax\arraymode\@sharp$\hfil
     \or $\relax\arraymode\@sharp$\hfil
     \or \hfil$\relax\arraymode\@sharp$\fi}}
\def\@array[#1]#2{\setbox\@arstrutbox=\hbox{\vrule
     height\arraystretch \ht\strutbox
     depth\arraystretch \dp\strutbox
     width\z@}\@mkpream{#2}\edef\@preamble{\halign
\noexpand\@halignto
\bgroup \tabskip\z@ \@arstrut \@preamble \tabskip\z@ \cr}%
\let\@startpbox\@@startpbox \let\@endpbox\@@endpbox
  \if #1t\vtop \else \if#1b\vbox \else \vcenter \fi\fi
  \bgroup \let\par\relax
  \let\@sharp##\let\protect\relax
  \@arrayskip\@preamble}
%
%
%
%
\def\eqnarray{\stepcounter{equation}%
              \let\@currentlabel=\theequation
              \global\@eqnswtrue
              \global\@eqcnt\z@
              \tabskip\@centering
              \let\\=\@eqncr
              $$%
 \halign to \displaywidth\bgroup
    \eqnumphantom\@eqnsel\hskip\@centering
    $\displaystyle \tabskip\z@ {##}$%
    \global\@eqcnt\@ne \hskip 2\arraycolsep
         $\displaystyle\arraymode{##}$\hfil
    \global\@eqcnt\tw@ \hskip 2\arraycolsep
         $\displaystyle\tabskip\z@{##}$\hfil
         \tabskip\@centering
    &{##}\tabskip\z@\cr}


\setcounter{footnote}0
\begin{center}
\hfill ITEP/TH-56/99\\
\vspace{0.3in}
{\LARGE\bf Dualities in integrable systems and
N=2 SUSY theories}
\date{today}

\bigskip {\Large A.Gorsky \footnote{
Talk given at QFTHEP-99, Moscow, May 29 - June 2 }}
\\
\bigskip
ITEP, Moscow, 117259, B.Cheryomushkinskaya 25\\

\end{center}
\bigskip

\begin{abstract}
We discuss dualities  of the integrable dynamics behind the
exact solution to the N=2 SUSY YM theory. It is shown that
T duality in the string theory is related to the separation
of variables procedure in dynamical system. We argue
that the are analogues of S duality as well as 3d mirror
symmetry in the many-body systems of Hitchin type governing
low-energy effective actions.
\end{abstract}

\section{Integrability and low-energy effective actions
for N=2 SUSY gauge theories}

The description of the strong coupling regime in
the quantum field theory remains a
challenging problem and the main hope is connected to discovery of
new proper degrees
of freedom which would provide the perturbative expansion
distinct from the initial
one. The first successful derivation of the low energy
effective action in N=2
SUSY Yang-Mills theory clearly shows
that solution of the theory involves new
ingredients which are not familiar
in this context before like
Riemann surfaces and meromorphic differentials on it \cite{sw}.

The general structure of the effective actions
is defined by the symmetry arguments,
in particular they should respect the Ward
identities coming from the bare field theory.
For example the chiral symmetry  fixes the chiral
Lagrangian in QCD and the conformal symmetry  provides
the dilaton effective
actions in N=0 and N=1 YM theories.
Since the effective actions have a symmetry
origin one can expect universality properties and generically different
UV theories can flow to the same IR ones.
It is the symmetry origin of the effective
actions that leads to  the appearance
of the integrable systems on the scene. The point is that the phase spaces
for the integrable systems coincide with some moduli space or the cotangent
bundle to the moduli space. We can mention  KdV hierarchy related to the
moduli of the complex structures of the Riemann surfaces, the Toda lattice
related to the moduli of the flat connections or Hitchin like systems
connected with the moduli of the holomorphic vector bundles. In any case
moduli spaces come from some additional symmetry of the problem.

Identification of the variables in the integrable system responsible for
some effective action is a complicated problem. At a moment there is no
universal way to introduce the proper variables in the theories which
are not topological ones but there is some experience in 2d theories
\cite{vafacec} which suggests to identify the nonperturbative
transition amplitudes among the vacuum states
as the dynamical variables.
As for the "space-time" variables, coupling
constants and sources are the most promising candidates. It is
expected that the partition
function evaluated in the low-energy effective
theory is the so called $\tau$ - function
in integrable hierarchy which
is the generating function for the conserved integrals of motion.
The particular solution of the
equations of motion in the dynamical system
is selected by applying the Ward identities to the partition function of
the effective theory.

The arguments above explain the reason for the search of some integrable
structures behind the Seiberg-Witten solution to N=2 SUSY Yang-Mills
theory. This integrable structures
which capture the hidden symmetry structure
have been found in \cite{gkmmm} where it was shown that $A_{N_{c}}$
affine Toda chain governs the low energy effective action and BPS
spectrum of pure N=2 SYM theory. The generalization to the theories
with matter involves Calogero-Moser integrable system for the adjoint
matter \cite{dw}  and XXX spin chain
for the fundamental matter \cite{chains}.
In five dimensions relativistic Toda
chain appears to be relevant for the pure
gauge theory \cite{nikita} while
anisotropic XXZ chain for SQCD \cite{ggm2}.
At the next step
completely anisotropic XYZ chain has been suggested as a guide for
6d SQCD \cite{ggm2}
while the generalization to the group product case is described by
the higher spin magnets \cite{ggm1}.
The candidate system for 6d theory
with adjoint matter based on the geometry of  elliptically fibered K3
manifold was suggested in \cite{fgnr} (see also \cite{bmmm}.
Therefore
there are no doubts in the validity
of the mapping between effective low-energy effective theories and
integrable finite dimensional systems.

The list of correspondences between two seemingly different issues
looks as follows. The solution to the classical equation of motion
in the integrable system can be expressed in terms of higher genus
Riemann surface which can be mapped to the complex Liouville
tori of the dynamical system. It is this Riemann surface enters
Seiberg-Witten solution, and the meromorphic differential introduced
to formulate the solution coincides with the action differential
in the dynamical system in the separated variables. The Coulomb
moduli space in N=2 theories is identified with the space of the
integrals of motion in the dynamical system, for example
$Tr{\phi}^{2}$ where $\phi$ is the adjoint scalar field coincides
with  the
Hamiltonian for the periodical Toda system. The parameters of
the field theory like masses or $\Lambda_{QCD}$
determine the parameters and couplings in the integrable system.
For instance in SQCD fundamental
masses provide the local Casimirs in the
periodical spin chains.

In spite of a lot of supporting facts it is necessary to get more
transparent explanation of the origin of integrability in this
context. To this aim let us discuss the moduli spaces in the problem
at hands. Classically there is only Coulomb branch of the moduli
space in pure gauge theory so one can expect dynamical system associated
with such phase space. Coulomb branch can be considered as a special
Kahler manifold  \cite{sw} while
the Hitchin like dynamical system responsible
for the model has a hyperkahler phase space
\cite{hitchin}. The resolution of the
contradiction comes from the hidden Higgs-like branch which has
purely nonperturbative nature
\cite{gorbr,ggm1}. It is the  dynamical system on this hidden
phase space provides the integrable system of the Hitchin
or spin chain type. Therefore
there are two moduli spaces in our problem and one expects a pair
of dynamical systems. This is what we have indeed; dynamical system
on the Higgs branch yields the Hitchin like dynamics with the
associated Riemann surfaces while the integrable system on the
Coulomb branch gives rise to the Whitham dynamics. The ``physical
meaning of the Hitchin system is to incorporate the nonperturbative
instanton like contributions to the
effective action in the supersymmetric
way while the Whitham dynamics is nothing
but the RG flows in the model \cite{gkmmm} \footnote{The latest
developments within Whitham approach as well as list of references
can be found in \cite{takasaki}}.

The next evident question is about the degrees of freedom in
both dynamical systems. The claim is that all degrees of the freedom can
be identified with the collective coordinates of a particular brane
configuration. First let us explain where the Higgs branch comes
from. The basic illustrative example for the
derivation of the hyperKahler moduli
space in terms of branes is the description of ADHM data
as a moduli for a  system  of coupled D1-D5
or D0-D4 branes \cite{doug}. If the gauge fields are
independent on some  dimension
one derives Nahm description of the monopole moduli space in terms
of D1-D3 branes configuration \cite{diac}. The transition from ADHM
data to the Nahm ones can be treated as a T duality transformation.
At the next step the  hyperkahler Hitchin space can be obtained by
reducing the dependence (or additional T duality transformation)
on one more dimension. This corresponds to  the system of D2 branes
wrapped around some surface $\Sigma$ holomorphically embedded in some
manifold. The most relevant example concerns
$T^{2}$  embedded into K3 manifold \cite{vafatop}. The T duality along
the torus transforms it to the system of D0 branes on the dual
torus, which is the most close picture for the Toda dynamics in terms
of D0 branes. The related discussion for the derivation of the
Hitchin spaces in terms of instantons
on $R^{2}\times T^{2}$ can be found in \cite{kap}.

Let us now proceed to the explicit brane picture for the
N=2 theories. There are different ways to get it, one involves
10d string theory which compactified on the manifold containing
the Toda chain spectral curve \cite{vafa}, or the M theory with
M5 brane wrapped around the noncompact surface which can be
obtained from the spectral curve by deleting the finite
number of points \cite{wittenm}. This picture can be considered as the
perturbative one  and nonperturbative degrees of freedom have to be
added. For this purpose it is useful to consider IIA projection of
the M theory which involves $N_{c}$ D4 branes between two NS5 branes
located on a distance $\frac{l_{s}}{g^{2}}$ along, say $x_{6}$
direction. Field theory is defined on
D4 branes worldvolume \cite{witbr} and
the extensive review concerning
the derivation of the field theories from branes
can be found in \cite{givkut}. The additional ingredient yielding
the hidden Higgs branch comes from the set of $N_{c}$ D0 branes,
one per each D4 brane \cite{gorbr,ggm1}. It is
known that D0 on D4 brane behaves
as a abelian point-like instanton but now we have the system of
interacting D0 branes. The coupling constant is provided by the
$\Lambda_{QCD}$ parameter which can be most naturally obtained
from the mass of the adjoint scalar breaking N=4 to N=2 via
dimensional transmutation procedure.

One way to explain the need for the additional D0,s
in IIA theory or KK modes in M theory looks as follows.
It is known that any finite-dimensional integrable system
with the spectral parameter allows the canonical transformation
to the variables -- spectral
curve with the linear bundle.
The spectral curve role is transparent and KK modes provide
the linear bundle. As we have already noted they are responsible
for the nonperturbative contribution but the summation of the
infinite instanton sums into the finite number degrees of freedom
remains the challenging problem.
It is worth noting that both canonical coordinates in the dynamical
system come from the coordinates of D0 branes in different
dimensions.
The necessity for the additional
nonperturbative degrees of freedom has been also discussed in
\cite{do}.

To show how the objects familiar in the integrability world translate
into the brane language consider two examples.
First let us consider the equations of motion in the Toda chain
which has the Lax form
\beq
\frac{dT}{ds}=[T,A]
\eeq
with some $N_{c}\times N_{c}$ matrixes T and A. The Lax matrix T
can be related to Nahm matrix for the chain of monopoles using
the identifications of the spectral curves for cyclic monopole
configuration and periodic Toda chain \cite{sut}.
All these
results in the following expression
for the Toda Lax operator in terms of the Nahm matrixes $T_{i}$
\be
T=T_{1}+iT_{2}-2iT_{3}{\rho}+(T_{1}-iT_{2}){\rho}^{2}
\ee
\be
T_{1}=\frac{i}{2}\sum_{j=1}q_{j}(E_{+j}+E_{-j})  \\
T_{2}=-\sum_{j=1}q_{j}(E_{+j}-E_{-j})             \\
T_{3}=\frac{i}{2}\sum_{j}p_{j}H_{j} ,
\ee
where E and H are the standard
SU(N) generators, $p_{i},q_{i}$ represent
the Toda phase space, and $\rho$
is the coordinate on the $CP^{1}$ above.
This $CP^{1}$ is involved in the twistor construction for
monopoles and a point on $CP^{1}$
defines the complex structure on the
monopole moduli space.
With these definitions Toda equation of motion and Nahm
equation acquire the simple form
\be
\frac{dT}{dt}=[T,A]
\ee
with fixed A.
Having in mind the brane interpretation of the
Nahm data \cite{diac} we can claim that the equations of motion provide
the conditions for the required supersymmetry of the whole configuration.

As another example of the validity of the brane-integrability
correspondence mention the possibility to incorporate the fundamental
matter in the gauge theory via branes in two ways. The first one
concerns the semiinfinite D4 branes while the second one the
set of $N_{f}$ D6 branes. One can
expect two different integrable systems behind and they were found
in \cite{kripho} and \cite{chains}. It was shown in \cite{ggm1} that
they perfectly correspond to the brane
pictures  and it appears that the
equivalence of two representations
agrees with some duality property in
the dynamical system. Recently one
more ingredient of the brane approach -
orientifold has been recognized within integrability approach \cite{gm}.
To conclude the discussion of the many-body
dynamical systems let us mention that one can inverse the logic and
use the possible integrable deformations of the dynamical system
to construct their field theory counterparts. Along this line of
reasoning we can expect some unusual field theories with the several
$\Lambda$  type scales \cite{ggm1}.

\section{Dualities in integrable systems and SYM theories}
We are going to study the phenomenon of {\it duality} whose
precise
definition is presented shortly. Duality is a subject of much recent
investigation in the context of
(supersymmetric) gauge theories,
in which case the duality is an
involution, which maps the observables
of one theory to those of another.
The duality is powerful  when the coupling
constant
in one theory
is inverse  of that in another (or more generally, when small coupling
is mapped to the strong one).
 For example,
a weakly coupled (magnetic) theory can be dual to the strongly coupled
(electric) theory thus making possible to understand the strong coupling
behavior
of the latter. In particular, it was shown \cite{sw}
that using the concept of duality one can find  exact low-energy
Lagrangian of $\CN=2$,$d=4$ $SU(2)$ gauge theory.
A more fascinating recent development is that the duality connecting
weak and strong coupling regimes of one or different theories
may have a geometric origin. The most notorious example of that
is provided by
$M$-theory. Having in mind the relation between many-body systems
and effective actions in SYM theories it is natural to
obtain the natural dualities within the integrability approach.
Both dynamical
systems and gauge theories benefit from establishing of this
correspondence which was formulated in \cite{gnr,fgnr}.
Brane picture for the
Hitchin like systems presented above
plays important role in derivation of
the proper degrees of freedom.

\subsection{T duality and separation of variables}
There are three essentially different dualities which
manifest themselves in dynamical systems of the
Hitchin type. Let us
start with the analogue of T duality in the Hitchin like
systems \cite{gnr}. It appears that the proper analogue
of T duality can be identified with the separation of variables
in the dynamical systems.

A way of  solving a
problem
with many degrees
of freedom is to reduce it to the problem with the
smaller
number of
degrees
of freedom.
The solvable models  allow to reduce the original system
with $N$
degrees
of freedom to $N$ systems with $1$ degree of freedom
which reduce to
quadratures. This approach is called a separation of
variables ( SoV). Recently,
E.~ Sklyanin formulated ``magic recipe'' for the  SoV in the
large
class of quantum integrable models with a Lax representation
\cite{sklyanin}. The method reduces in the classical case to the
technique of separation of variables using poles of the Baker-Akhiezer
function (see also \cite{kripho})
for recent developments and more references). The
basic
strategy of this method is to look
at the Lax   eigen-vector ( which is the  Baker-Akhiezer
function) $\Psi (z, \lambda)$:
\beq
L(z) \Psi(z, \lambda) = \lambda (z) \Psi(z,
\lambda)
\eeq
with some choice of normalization.
The poles $z_{i}$ of $\Psi(z, \lambda)$
together with the eigenvalues
$\lambda_{i} = \lambda(z_{i})$ are the
separated variables. In all the examples
studied so far the most naive way of
normalization leads to the canonically
conjugate coordinates $\lambda_{i},
z_{i}$.

Remind that the phase space for the  Hitchin system can be
identified with the cotangent bundle to the moduli space of
holomorphic vector  bundle $T^{*}M$ on the surface $\Sigma$.
The following symplectomorphysms can be identified with
the separation of variables procedure. The phase space
above allows two more formulations; as the pair $(C,\cal{L})$
where C is the spectral curve of the dynamical system and
$\cal{L}$ is the linear bundle or  as the Hilbert scheme
of points on $T^{*}\Sigma $ where the number of points follows
from the rank of the gauge group. It is the last
formulation provides the separated variables. The
role of Hilbert schemes on   $T^{*}\Sigma $   in context
of Hitchin system was established for the surfaces
without marked points in \cite{hurt} and generalized
for the systems of Calogero types in \cite{wilson, naka, gnr}.

In the brane terms separation of variables can be
formulated  as reduction to a system of D0 branes
on some four dimensional manifold. It reminds
a reduction to a system of point-like instantons
on a (generically noncommutative \cite{ns}) four manifold.
One more essential point is that separated variables
amount to some explanation of the relation of periodic
Toda chain above   and monopole chains. Indeed, monopole
moduli space have the structure resembling the
one for the Toda chain in separated variables; both
of them are the Hilbert schemes of points on the
similar four manifolds.

The abovementioned constructions of the separation of variables
in integrable systems on moduli spaces of holomorphic bundles
with some additional structures can be described
as a symplectomorphism between  the moduli spaces of
the bundles (more precisely, torsion free sheaves)
having different Chern classes.

To be specific let us concentrate on the moduli space
$\CM_{\vec v}$ of stable torsion free coherent sheaves ${\CE}$
on $S$. Let ${\hat A}_{S} = 1 - [ {\rm pt} ] \in H^{*} (S, Z)$
be the $A$-roof
genus of $S$. The vector
$\vec v = Ch ({\CE}) \sqrt{\hat A_{S}} =
(r; \vec w; d - r) \in {H}^{*}(S, Z), \vec w \in \Gamma^{3,19}$
corresponds to the sheaves
 with the Chern numbers:
\beq
ch_{0} ({\CE})  = r \in {H}^{0}(S ; Z)
\eeq
\beq
ch_{1}({\CE})  = \vec w \in {H}^{2} (S; Z)
\eeq
\beq
ch_{2}({\CE})  = d \in {H}^{4}(S; Z)
\eeq
Type $II A$ string theory compactified on $S$ has BPS
states, corresponding to
the $Dp$-branes, with $p$ even, wrapping
various supersymmetric cycles in $S$, labelled by
$\vec v \in {H}^{*}(S, Z)$. The actual states
correspond to the cohomology classes of the moduli spaces
$\CM_{\vec v}$
of the configurations of branes.  The latter can be identified
with the moduli spaces $\CM_{\vec v}$ of appropriate sheaves.

The string theory, compactified on $S$ has moduli space
of vacua, which can be identified with
$$
\CM_{A} = O\left( {\Gamma}^{4,20} \right) \backslash
O(4,20; R) / O(4;R) \times O(20;R)
$$
where the arithmetic group $O({\Gamma}^{4,20})$ is the group
of discrete authomorphismes. It maps the states
corresponding to different $\vec v$ to each other.
The only invariant of its action is ${\vec v}^{2}$.

We have studied three realizations of an
integrable system.
The first one uses the non-abelian
gauge fields on the
curve $\Sigma$ imbedded into symplectic
surface $S$. Namely,  the phase space of the system is the
moduli space of stable pairs: $(\CE, \phi)$, where $\CE$
is rank $r$ vector bundle over $\Sigma$ of degree
$l$, while $\phi$
is the holomorphic section of
$\omega^{1}_{\Sigma} \otimes {\rm End}({\CE})$.
The second realization is the moduli space of pairs $(C, {\CL})$,
where $C$ is the curve (divisor) in $S$ which realizes
the homology class $r[\Sigma]$ and $\CL$ is the line bundle on $C$.
The third realization is the Hilbert scheme of points on $S$
of length $h$, where $h = {\half}{\rm dim}{\CM}$.

The equivalence of the first and the second realizations
corresponds to the physical statement that the bound
states of $N$ $D2$-branes wrapped around $\Sigma$ are represented
by a single $D2$-brane which wraps a holomorphic curve $C$
which
is an $N$-sheeted covering of the base curve $\Sigma$.
The equivalence of the second and the third descriptions
is natural to attribute to $T$-duality.

Let us mention that the separation of variables
above provides some insights on the Langlands duality
which involves spectrum of the Hitchin Hamiltonians.
The attempt to reformulate Langlands duality as a quantum
separation of variables has been successful for the
Gaudin system corresponding to the spherical case \cite{frenkel}.
The consideration in \cite{fgnr}  suggests that the  proper
classical version of the Langlands correspondence
is the transition to the Hilbert scheme of points
on four-dimensional manifold. This viewpoint
implies that quantum case can be considered as
correspondence   between
the eigenfunctions of the Hitchin Hamiltonians
and solutions to the Baxter equation in the separated
variables.

\subsection{S-duality}

Now let us explain that S-duality
well established in the field theory  also
has clear counterpart
in the holomorphic dynamical system.

The action variables in dynamical system
are the integrals of meromorphic differential $\lambda$ over
the $A$-cycles on the spectral curve.
The reason for the $B$-cycles to be
discarded is simply the fact that the $B$-periods of $\lambda$
are not independent of the $A$-periods. On the other
hand, one can choose as the independent periods the
integrals of $\lambda$ over any lagrangian
subspace in $H_{1}({\bf T}_{b}; Z)$.

This leads to the following structure of the action variables
in the holomorphic setting. Locally over a disc in $B$ one
chooses a basis in $H_{1}$ of the fiber together with the
set of $A$-cycles. This choice may differ over another
disc. Over the intersection of these discs one has
a $Sp(2m, Z)$ transformation relating the bases. Altogether
they form an $Sp(2m, Z)$ bundle. It is an easy excercise on the
properties of the period matrix that the two form:
\beq
dI^{i} \wedge dI^{D}_{i}
\eeq
vanishes. Therefore one can always locally  find a function
$\CF$ - {\it prepotential}, such that:
\beq
I^{D}_{i} = {{\p \CF}\over{\p I^{i}}}
\eeq
The angle variables are uniquely reconstructed once the action
variables are known.

To illustrate the meaning of the action-action
{\sl AA} duality we look at the
two-body system, relevant for the
$SU(2)$ $\CN=2$ supersymmetric
gauge theory :
\beq
H =
{{p^{2}}\over{2}} + \Lambda^{2} \cos (q)
\eeq
with $\Lambda^{2}$ being a complex
number - the coupling constant of a two-body
problem and at the same time a
dynamically generated scale of the gauge theory.
The action variable is
given by one of the periods of the differential $pdq$.
Let us introduce more
notations:
$x = \cos (q)$, $y = {{p\sin (q)}\over{\sqrt{-2}\Lambda}}$, $u =
{{H}\over{\Lambda^{2}}}$. Then the
spectral curve, associated to the system
which is also a level set of the Hamiltonian
can be written as
follows:
\beq
y^{2} = (x - u) (x^{2}-1)
\eeq
which is exactly
Seiberg-Witten curve .
The periods are:
\beq
I = \int_{-1}^{1} \sqrt{{x-u}\over{x^{2}-1}} {dx},
I^{D} = \int_{1}^{u} \sqrt{{x-u}\over{x^{2}-1}} {dx}
\eeq
They obey  Picard-Fuchs equation:
$$
\left( {{d^{2}}\over{du^{2}}} +
 {1\over{4(u^{2}-1)}} \right) \pmatrix{& I\cr& I^{D}\cr} = 0
$$
which can be
used to write down an asymptotic expansion of the action
variable near
$u=\infty$ or $u= \pm 1$ as well as that of prepotential . The {\sl
AA} duality is manifested in the fact that
near $u =\infty$ (which
corresponds to the high energy scattering in the two-body problem
and also a
perturbative regime of $SU(2)$ gauge theory) the appropriate action
variable
is $I$ (it experiences a monodromy $I \to - I$ as $u$ goes around
$\infty$), while
near $u = 1$ (which corresponds to the dynamics of the
two-body
system near the top of the potential and to the strongly coupled
$SU(2)$ gauge theory)
the appropriate variable is $I^{D}$ (which corresponds
to a weakly coupled magnetic $U(1)$ gauge theory and is actually well
defined near $u=1$ point) . The monodromy
invariant combination of the
periods \cite{matone}:
\beq
II^{D} - 2\CF = u
\eeq
(whose origin is in the periods of
Calabi-Yau manifolds on the one hand
and in the  properties of anomaly in theory on the other) can be chosen
as a global coordinate on the space of integrals of motion .
At $u \to \infty$ the prepotential has an expansion of the form:
$$
\CF \sim {\half}  u \log u + \ldots \sim I^{2}
\log I + \sum_{n}{{f_{n}}\over{n}} I^{2-4n}
$$

Let us emphasize that S-duality maps the dynamical
system to itself. We have seen that the notion of prepotential can be
introduces for any holomorphic many-body system
however its physical meaning  as well as its properties
deserve further investigation.

\subsection{"Mirror" symmetry  in dynamical systems}

The last type of duality we would like to discuss concerns
dualities between pair of dynamical systems \cite{ru,fgnr}.
To start with let us remind how this symmetry is formulated
within the field theory. The initial motivation amounts
from the 3d theory example \cite{is} where mirror symmetry
interchanges Coulomb and Higgs branches of the moduli space.
The specifics of three dimensions is that both Coulomb and
Higgs branches are hyperkahler manifolds and the mirror
symmetry can be formulated as a kind of hyperkahler rotation.
The attempt to formulate the similar symmetry for 4d theory
was performed in \cite{kapustin}.

In \cite{fgnr} the general procedure for the analogous
symmetry within integrable system
in terms of Hamiltonian and Poissonian
reductions was formulated. Symmetry
maps one  dynamical system with coordinates $x_i$    to
another one whose coordinates coincide with the action
variables of the initial system and vise versa. It appears
that taking into account the relation between dynamical
systems and low-energy effective actions this duality
in general
maps Higgs and Coulomb branches of the moduli space
in the  gauge
theories in different dimensions.

Qualitatively this symmetry is even more
transparent in terms
of separated variables. As was discussed above
the proper object in separated variables
is the hyperkahler
four dimensional manifold
which provides  the phase space. In most
general situation the manifold involves two tori  or
elliptically fibered K3 manifold.
One torus provides momenta while
the seconds coordinates. The duality at hands
actually interchanges momentum and coordinate tori
and in generic case self-duality is expected. Corresponding
field theory counterpart is the hypothetical six-dimensional
theory with adjoint matter.

All other cases correspond to some degeneration. Degeneration
of the momentum torus to $C/Z_{2}$ corresponds to the transition
to the five-dimensional theory while degeneration  to $R^2$
corresponds  to four-dimensional theory. Since the modulus
of the coordinate torus  has the meaning of the
complexified bare coupling in the theory the interpretation
of the degeneration of the coordinate torus is different.
Degeneration to the cylinder corresponds to the switching
off the instanton effects while the rational degeneration
corresponds to the additional degeneration.

We will see below that the "mirror" symmetry maps theories
in different dimensions to each other. Instanton effects
in one theory "map" into the additional compact dimension
in the dual counterpart. We will discuss mainly classical
case with only few comments on the quantum picture. Since
the wave functions in the Hitchin like systems can be
identified with some solutions to the KZ or qKZ equations
the quantum duality would mean some relation between
solutions to the rational, trigonometric or elliptic KZ
equations. Recently the proper symmetries for KZ equations
where discussed in \cite{etingof}.

\subsubsection{Two-body system (SU(2))}
Let us discuss first two-body system corresponding to SU(2) case.
Two-particle systems which we are going to consider  reduce (after
exclusion of the center of mass motion) to a one-dimensional problem.
The action-angle variables can  be written explicitly  and
the
dual system emerges immediately once the natural Hamiltonians
are chosen. The problem is the following. Suppose
the phase space is coordinatized by $(p,q)$. The dual Hamiltonian
(in the sense of AC duality) is a function of $q$
expressed in terms of $I, \varphi$, where $I, \varphi$ are the action-angle
variables of the original system :  $H_{D}(I, \varphi) = H_{D}(q)$. In all
the cases  below there is
a natural choice of $H_{D}(q)$.

Consider as example elliptic Calogero model whose Hamiltonian is:
\beq
H(p,q) = {{p^{2}}\over{2}}  + {\nu^{2}} {\wwp}.
\eeq
Here $p,q$ are complex, $\wwp$ is the Weierstrass
function on the elliptic curve $E_{\tau}$:
\beq
{\wwp} = {1\over{q^{2}}} + \sum_{\matrix{&(m,n) \in
{Z^{2}} \cr  &(m,n) \neq  (0,0) \cr}}
{1\over{(q + m\pi  +n \tau\pi )^{2}}} - {1\over{(m\pi  +n \tau\pi)^{
2}}}
\eeq
Let us introduce the Weierstrass notations:
$x = {\wwp}$, $y = {\wwp}^{\prime}$.
We have an equation defining the curve $E_{\tau}$:
\beq
y^{2} = 4x^{3} - g_{2}(\tau) x -  g_{3}(\tau) =
4 \prod_{i=1}^{3} (x - e_{i}), \quad \sum_{i=1}^{3} e_{i} = 0
\eeq
The holomorphic differential $dq$ on $E_{\tau}$ equals $dq = dx/y$.
Introduce the variable $e_{0} = 2E/{\nu}^{2}$.
The action variable is one of the periods of the
differential ${pdq}\over{2\pi}$ on the curve
$E = H(p,q)$ :
\beq
I = {1\over{2\pi}} \oint_{A} \sqrt{2(E - {\nu}^{2} {\wwp}) } =
{1\over{4\pi i}} \oint_{A}
{{dx \sqrt{x - e_{0}}}\over{\sqrt{(x-e_{1})(x-e_{2})(x-e_{3})}}}
\eeq
The angle variable can be determined from the condition
$dp \wedge dq = dI \wedge d\varphi$:
\beq
d\varphi = {1\over{2i T(E)}} {{dx}\over
{\sqrt{\prod_{i=0}^{3}(x - e_{i})}}}
\eeq
where $T(E)$ normalizes $d\varphi$ in such a way that
the  $A$ period of $d\varphi$ is
equal to $2\pi$:
\beq
T(E) ={1\over{4\pi i }}
\oint_{A} {{dx}\over{\sqrt{\prod_{i=0}^{3} (x - e_{i})}}}
\eeq
Thus:
\beq
2i T(E)
d\varphi =  {{dx}\over{\sqrt{4 \prod_{i=0}^{3}
(x - e_{i})}}}
\eeq
\beq
\omega d\varphi =
{{dt}\over{\sqrt{4 \prod_{i=1}^{3} ( t- t_{i})}}}
\eeq
where
\beq
\omega = - 2i T(E) \sqrt{e_{01}e_{02}e_{03}}   =
{1\over{2\pi}} \oint_{A}  {{dt}\over{\sqrt{4 \prod_{i=1}^{3}
( t- t_{i})}}}
\eeq

\beq
t=  {1\over{x-e_{0}}} + {1\over{3}}
\sum_{i=1}^{3} {1\over{e_{0i}}} ; t_{i} = {1\over{3}}
\sum_{j=1}^{3} {{e_{ji}}\over{e_{0i} e_{0j}}}
\eeq
where $e_{ij} =  e_{i} - e_{j}$

Introduce a meromorphic function on $E_{\tau}$:
\beq
{\widehat {cn}}_{\tau}(z) = \sqrt{{x-e_{1}}\over{x-e_{3}}}
\eeq
where $z$ has periods $2\pi$ and $2\pi \tau$.
It is an elliptic analogue of the cosine (in fact, up to a rescaling
of $z$ it coincides with the Jacobi elliptic cosine).
Then we have:
\beq
H_{D}(I, \varphi) = {\widehat {cn}}_{\tau}(z) =
{\widehat {cn}}_{\tau_{E}}({\varphi})
\sqrt{1 -{{{\nu}^{2}e_{13}}\over{2E - {\nu}^{2}e_{3}}}}
\eeq
where $\tau_{E}$ is the modular parameter of the relevant spectral curve
$v^{2} = 4 \prod_{i=1}^{3} ( t - t_{i} )$:
\beq
\tau_{E} = \Bigl( \oint_{B} {{dt}\over{\sqrt{4 \prod_{i=1}^{3}
(t - t_{i})}}}\Bigr) \Bigr/ \Bigl( \oint_{A}
{{dt}\over{\sqrt{4 \prod_{i=1}^{3}
(t - t_{i})}}} \Bigr).
\eeq
For large $I$, $2E(I) \sim I^{2}$.

Therefore  the elliptic Calogero
model with rational dependence on momentum and elliptic on
coordinate   maps into the "mirror" dual system with
elliptic dependence on momentum and rational on coordinate.
On the field theory side d=4 theory with adjoint matter
maps into the d=6 theory with adjoint matter with
the instanton corrections switched off. The coordinates
on the Coulomb branch in d=4 theory becomes the coordinates
on the "Higgs branch " in d=6 theory which explains the
origin of the term "mirror" symmetry in this context.

\subsubsection{Many-body systems}
Now we would like to demonstrate how the "mirror"
transform can be formulated in terms of Hamiltonian
or Poissonian reduction procedure. It appears that
it corresponds in some sence to the simultaneous
change of the gauge fixing and Hamiltonians. More
clear meaning of these words will be clear from the
examples below.

We summarize the systems
and their duals in rational
and trigonometric cases in the following table:
\beq
\matrix{& &{\rm rat. CM} &
\leftrightarrow & {\rm rat. CM} & &\cr
& R \to 0 &   \uparrow  &
                 &  \uparrow       & \beta \to 0& \cr
& &{\rm trig. CM} & \leftrightarrow & {\rm rat. RS} & &\cr
& \beta \to 0 &   \uparrow  &
                        &  \uparrow       &R \to 0&\cr
& &{\rm trig. RS} & \leftrightarrow & {\rm tri
g. RS} & &\cr}
\eeq

Here $CM$ denotes ${\it Calogero-Moser}$ models  and
$RS$ stands for ${\it Ruijsenaars-Schneider}$.
The parameters $R$ and $\beta$ here are the radius of the circle
the coordinates of the particles take values in and the
inverse speed
of light respectively.
The horizontal arrows in this table are the dualities, relating
the systems on the both sides. Most of them were discussed by
Simon Ruijsenaars
\cite{ru}. We
notice that the duality transformations form a
group which  in the case of self-dual systems listed here contains
${\rm SL}_{2}(Z)$.
The generator $S$ is the gorizontal arrow described below, while the
$T$ generator is in fact a certain finite time evolution of the
original system (which is always
a symplectomorphism, which maps
the integrable system to the dual one).
\noindent
We begin with recalling the Hamiltonians of these systems.
Throughout this section
$q_{ij}$ denotes $q_{i} - q_{j}$.

Consider the space
 ${\CA}_{{\bf T}^{2}}$ of
$SU(N)$ gauge fields $A$ on a two-torus
${\bf T}^{2} = {\bf S}^{1} \times {\bf S}^{1}$.
Let the circumferences of the circles
be $R$
and $\beta$.
The space  ${\CA}_{{\bf T}^{2}}$ is acted on by a gauge group $\CG$ ,
which preserves a symplectic form
\beq
\Omega = {k\over{4{\pi}^{2}}}
\int {\Tr} \delta A \wedge \delta A,
\eeq
with $k$ being an arbitrary real
number for now.
The gauge group acts via evaluation at
some point $p \in {\bf T}^{2}$
on any coadjoint orbit $\CO$ of $G$, in particular, on $\CO =
{\IC\IP}^{N-1}$.
Let $(e_{1} :  \dots : e_{N})$
be the homogeneous
coordinates on $\CO$. Then the moment map for the action of $\CG$
on $\CA_{{\bf T}^{2}} \times \CO$ is
\beq
k F_{A} + J \delta^{2}(p),  \quad J_{ij}
= i{\nu} (  \delta_{ij} - e_{i} e_{j}^{*})
\eeq
$F_{A}$ being the curvature
two-form. Here we think of $e_{i}$ as
being the coordinates on $\IC^{N}$
constrained so that $\sum_{i} \vert e_{i} \vert^2  =N$ and
considered up to the multiplication
by a common phase factor.

Let us
provide a certain amount of commuting Hamiltonians. Obviously,
the
eigen-values of the monodromy of $A$
along any fixed loop on ${\bf T}^{2}$
commute
with themselves. We consider the
reduction at the zero level of the moment
map.  We have at least $N-1$ functionally independent commuting
functions on
the reduced phase space $\CM_{\nu}$.

Let us estimate the dimension
of
$\CM_{\nu}$.
If $\nu= 0$ then the moment equation forces the connection to be
flat and therefore
its gauge orbits are parameterized by the conjugacy
classes of the monodromies
around two non-contractible cycles on ${\bf T}^{2}$: $A$
and $B$. Since the fundamental
group $\pi_{1} ({\bf T}^{2})$ of ${\bf T}^{2}$
is abelian $A$ and $B$ are to
commute. Hence they are simultaneously diagonalizable, which makes
${\CM}_{0}$ a
$2(N-1)$ dimensional manifold. Notice that the generic point on the
quotient
space has a nontrivial stabilizer, isomorphic to the maximal torus
$T
$ of $SU(N)$. Now, in the presence of $\CO$
the moment equation implies
that the connection $A$ is flat outside of $p$ and has a
nontrivial
monodromy around $p$. Thus:
\beq
ABA^{-1}B^{-1} = {\exp}(R\beta J)
\eeq
(the factor $R\beta$ comes from the normalization of the
delta-function ).
If we diagonalize
$A$, then $B$ is uniquely
reconstructed up to the right
multiplication by the elements of $T$. The
potential degrees of freedom in $J$ are
"eaten" up by the former stabilizer
$T$ of a flat connection: if we conjugate both
$A$ and
$B$ by an element $t
\in T$ then $J$ gets conjugated. Now, it is important that
$\CO$ has
dimension $2(N-1)$. The reduction of
$\CO$ with respect to $T$ consists of a
point and does not contribute to the
dimension of $\CM_{\nu}$. Thereby we
expect to get an integrable system.
Without doing any computations we already
know
that we get a pair of dual systems. Indeed, we may choose as the set of
coordinates the eigen-values
of $A$ or the eigen-values of $B$.

The two-dimensional
picture has the advantage that the geometry of the problem
suggest the
$SL_{2}(Z)$-like duality. Consider the operations $S$ and $T$  realized
as:
\beq
S: (A, B) \mapsto (ABA^{-1}, A^{-1}); \quad T: (A,B)
\mapsto (A,  BA)
\eeq
which correspond to the freedom of  choice  of generators  in
the fundamental group of a two-torus. Notice that
both $S$ and $T$ preserve the commutator $ABA^{-1} B^{-1}$ and
commute with the action of the gauge group.
The group $\Gamma$ generated by $S$ and $T$
in the limit $\beta, R \to 0 $
contracts to ${\rm SL}_{2}(Z)$ in a sense that
we get the
transformations
by expanding
$$
A = 1 + \beta P + \ldots, \quad B = 1 + R Q +
\ldots
$$
for $R, \beta \to 0$.

The disadvantage of the two-dimensional
picture us the necessity to keep too many redundant
degrees of freedom. The
first of the contractions  actually allows to replace the space of
two
dimensional gauge fields by the
cotangent space to the (central extension of)
loop group:
$$
T^{*}{\hat G}   = \{ ( g(x), k\p_{x} + P(x) ) \}
$$
which is a
``deformation'' of the phase space of the previous example
($Q(x)$ got
promoted to a group-valued field). The relation to the two dimensional
construction
is the following. Choose a non-contractible circle $\bf S^{1}$
on the two-torus
which does not pass through the marked point $p$. Let $x,y$
be the coordinates on the torus
and $y=0$ is the equation of the $\bf S^{1}$.
The periodicity  of $x$ is $\beta$ and that of $y$ is $R$.
Then
$$
P(x) = A_{x}(x,0),
g(x) =P\exp\int_{0}^{R} A_{y}(x,y) dy.
$$
The moment map equation looks as follows:
\beq
k g^{-1} \p_{x}
g + g^{-1}Pg - P = J \delta(x),
\eeq
with $k = {1\over{R\beta}}$. The solution of
this equation in the gauge $P = {\rm diag}(q_{1}, \ldots, q_{N})$
leads to
the Lax operator $A = g(0)$ with $R,\beta$
exchanged. On the other hand,
if we  diagonalize
$g(x)$:
\beq
g(x) =  {\rm diag} \left( z_{1} =
e^{iR q_{1}}, \ldots, z_{N} = e^{iR q_{N}} \right)
\eeq
then a similar calculation
leads to the Lax operator
$$
B = P\exp\oint{1\over{k}} P(x)dx =  {\rm diag} (
e^{i \theta_{i} } ) \exp iR\beta\nu {\rm r}
$$
with
$$
{\rm r}_{ij} =
{1\over{1- e^{iRq_{ji}}}}, i\neq j; \quad {\rm r}_{ii} = - \sum_{j\neq i}
{\rm r}_{ij}
$$
thereby establishing the duality $A \leftrightarrow B$
explicitly.

When Yang-Mills theory is formulated on a cylinder
with the insertion of an appropriate time-like Wilson line, it
is equivalent to the Sutherland model describing a collection
of $N$ particles on a circle. The observables ${\Tr} \phi^k$
are precisely the integrals of motion of
this system.
One can look at other supercharges as well. In particular,
when the theory is formulated on a cylinder there is another
class of observables annihilated
by a supercharge. One can arrange the
combination
of supercharges which  will annihilate the Wilson loop operator. By repeating
the procedure similar to the one in  one arrives at the
quantum mechanical theory whose Hamiltonians are generated
by the spatial Wilson loops. This model is nothing
but the rational Ruijsenaars-Schneider many-body system.

The self-duality of trigonometric Ruijsenaars system has even
more transparent physical meaning. Namely, the field theory
whose quantum mechanical avatar is the Ruijsenaars system is
three dimensional Chern-Simons theory on
${\bf T}^{2} \times {\bf R}^{1}$ with the insertion of an appropriate
temporal Wilson line and spatial Wilson loop. It is the freedom
to place the latter which leads to several equivalent theories.
The group of (self-)dualities of this model is very big
and is generated by the transformations $S$ and $T$ .

Finally let us comment on
six dimensional theory compactified on a three
dimensional torus ${\bf T}^{3}$ down to three dimensions.
As was discussed extensively in \cite{sw3}
in case where two out of three radii of ${\bf T}^{3}$ are much smaller
then the third one ${\bf R}$ the
effective three dimensional
theory is a sigma model with the target space ${\CX}$ being the
hyper-kahler manifold (in particular, holomorphic symplectic)
which is a total space of algebraic integrable system. The complex structure
in which ${\CX}$ is the algebraic integrable system
is independent of the radius ${\bf R}$ while the K\"ahler structure
depends on $\bf R$ in such a way that the K\"ahler class of the
abelian fiber is proportional to $1/{\bf R}$.

The theory in three dimensions which came from four dimensions
upon a compactification on a circle whose low-energy
effective action describes only abelian degrees of freedom
can be always dualized to the theory of scalars/spinors only,
due to the vector-scalar duality in three dimensions.
In this way different sets of vector and hypermultiplets
in four dimensions can lead to the same three dimensional
theory  \cite{is,kapustin}.

I am indebted to V. Fock, N.Nekrasov and V. Roubtsov
for the collaboration on this subject.
This work is supported in part by grant INTAS-97-0103.

\end{document}